\documentclass[aps,prd,10pt,twocolumn,showpacs,preprintnumbers,amsmath,amssymb,nofootinbib]{revtex4-1}
\usepackage{graphicx,color,mathtools}
\usepackage{dcolumn}
\usepackage{bm}
\usepackage[mathscr]{euscript} 

\usepackage{amsthm} 

\def\t{\tilde}

\def\crit{{\rm crit}}

\def\statvolt{{\rm statvolt}}

\begin{document}

\title{A Note on the Kaluza-Klein Theory} 

\author{Li-Xin Li}
\email{lxl@pku.edu.cn, prof.li.xin.li@gmail.com}
\affiliation{Kavli Institute for Astronomy and Astrophysics, Peking University, Beijing 100871, P. R. China}

\date{\today}

\begin{abstract}
  We show that the Kaluza-Klein theory contains a fundamental problem: The four-dimensional metric tensor and the electromagnetic potential vector assumed in the Kaluza-Klein theory belong to four-dimensional vector spaces that are not integrable in general, resulting that the four-dimensional physical variables and the corresponding field equations derived from the five-dimensional Einstein field equation (i.e., the four-dimensional Einstein field equation and the Maxwell equations) are not defined on a four-dimensional submanifold. That is, the four-dimensional spacetime assumed in the Kaluza-Klein theory does not exist. No satisfactory solutions are found within the Kaluza-Klein formalism. Perhaps the best approach to fix the problem is giving up the Kaluza-Klein theory and looking for a new unified scheme for gravitational and electromagnetic interactions in the framework of a spacetime with extra dimensions, as having already been explored in some literature.
\end{abstract}


\maketitle

\section{Introduction}

The Kaluza-Klein (KK) theory represents the first attempt to unify the gravitational and electromagnetic interactions in the framework of general relativity extended to a spacetime with extra dimensions \citep{kal21,kle26a,kle26b}.\footnote{An even earlier attempt to unify electromagnetic and gravitational fields in a five-dimensional spacetime before the appearance of general relativity was given by Nordstr{\"o}m in 1914 \cite{nor14}.} In the KK theory, the bulk spacetime is assumed to be five-dimensional and described by the five-dimensional Einstein field equation. By 4+1 decomposition of the five-dimensional spacetime metric, a four-dimensional Einstein field equation and the Maxwell equations are derived from the five-dimensional Einstein field equation, which are assumed to describe the four-dimensional world where we live. How to make the extra dimension compact, small, and static has been a challenging problem in modern theoretical physics \citep{cre76,cre77,luc78,eng82,li98}. Nowadays, introducing (compact or noncompact but warped) extra dimensions in addition to the four dimensions of the spacetime where we live has been a popular strategy for unifying all fundamental interactions in nature, e.g., the theories of supergravity \citep{nie81,duf86}, superstring \citep{gre12,pol03,bec07,kir19}, and brane gravity \citep{ran99a,ran99b,maa10}. To test the existence of extra dimensions, the KK particles arising from the excitation of fields along the compact extra dimensions have been extensively searched by the Large Hadron Collider \citep{nat99,bha09,deu17}.

Despite the success of the KK theory in derivation of the Maxwell equations from the higher-dimensional Einstein field equation and its heavy influence on modern theories of unification, in this paper we show that the KK theory has a serious problem in its foundation: The four-dimensional metric tensor and the electromagnetic potential vector assumed in the KK theory are defined in vector spaces that are not integrable hence not tangent to any four-dimensional submanifold, unless the electromagnetic field antisymmetric tensor vanishes. That is, the four-dimensional spacetime assumed in the KK theory to support the effective four-dimensional theory of electromagnetism and gravity is defined only when the electromagnetic field vanishes, which conflicts the original aim of the KK theory in unifying the gravitational and electromagnetic interactions.

The influence of the problem just mentioned may not be limited to the KK theory. As is well known, one of the cornerstones of string theory---extra dimensions and compactification of extra dimensions---originated from the KK theory with an extension from one extra dimension to multiple extra dimensions. In string theory, a popular approach to derive gauge fields from higher-dimensional gravity is through the KK mechanism with an extension to spacetime of dimensions greater than five. For example, this is the case in the eleven-dimensional supergravity when it is connected to the low-energy limit of M-theory \citep{pol03,bec07,kir19}.

The paper is organized as follows. In section \ref{KK}, we outline the KK theory and derive the representation of the four-dimensional metric tensor and the electromagnetic potential vector in the five-dimensional spacetime. The representation is uniquely determined by the self-consistency requirement of the theory. In section \ref{geom}, we discuss the geometric interpretation of the above two KK variables and quantities derived from them (e.g., the electromagnetic field antisymmetric tensor). We show that the four-dimensional quantities are in vector spaces orthogonal to the direction of the extra dimension. In section \ref{subm}, we prove that the vector spaces containing the four-dimensional variables are not integrable unless the electromagnetic field tensor vanishes. Thus, in general there does not exist a four-dimensional submanifold supporting the four-dimensional theory derived from the five-dimensional Einstein field equation.

Section \ref{action} is devoted to discussion on the action principle and compactification of the extra dimension under the assumption of the cylinder condition. We show that after compactification, although the four-dimensional Einstein field equation and the Maxwell equations can be derived from the action principle, there still does not exist a four-dimensional submanifold supporting the four-dimensional field equations. Finally, in section \ref{disc}, we summarize the results that we have obtained in this paper and discuss their implications.

Throughout the paper geometrized units with $G=c=1$ are adopted unless otherwise stated, where $G$ is the four-dimensional gravitational constant and $c$ is the speed of light.  In addition, we will take $(-,+,+,+,+)$ as the convention for the signature of the five-dimensional spacetime metric. The abstract index notation for vectors and tensors advocated in \cite{wal84} will be used. That is, vectors and tensors are denoted by letters followed by lower case Latin indices, e.g., $v^a$, $g_{ab}$, etc.

\section{The Kaluza-Klein Formalism}
\label{KK}

The success of the KK theory relies on a specific decomposition scheme of the metric tensor of a five-dimensional bulk spacetime. Without loss of generality, in a five-dimensional spacetime $(\t{\cal M},\t{g}_{ab})$ we take a coordinate system $\{x^0,x^1,x^2,x^3,x^4=w\}$ and write the matrix representation of the five-dimensional metric tensor $\t{g}_{ab}$ as\footnote{The form of metric decomposition in equation (\ref{tg_ab}) agrees with the general case for the KK theory generalized to a $(4+n)$-dimensional spacetime to include non-Abelian gauge fields, where the $\phi^2$ is replaced by a matrix $g_{ij}$ with the indices $i$ and $j$ running from $1$ to $n$ for the extra dimensions \citep{dew64,ker68,cho75}.}
\begin{eqnarray}
  \t{g}_{AB} = \left(\begin{array}{cc}
    g_{\mu\nu} +\phi^2A_\mu A_\nu & \phi^2 A_\mu \\[1mm]
    \phi^2 A_\nu & \phi^2
  \end{array}\right) \;, \label{tg_ab}
\end{eqnarray}
where indices $A, B=0,1,2,3,4$, and $\mu,\nu=0,1,2,3$. Capital Latin letters label coordinate components of five-dimensional vectors and tensors. Lower case Greek letters label coordinate components of four-dimensional vectors and tensors.

The $4\times 4$ matrix $g_{\mu\nu}$ is interpreted as the component of the metric on a four-dimensional spacetime $({\cal M},g_{ab}$) associated with the coordinate system $\{x^0,x^1,x^2,x^3\}$, the $4\times 1$ matrix $A_\mu$ as the component of an electromagnetic potential dual vector, and the function $\phi$ as a scalar field in $({\cal M}, g_{ab})$. With the convention in equation (\ref{tg_ab}), the five-dimensional spacetime metric tensor $\t{g}_{ab}$ is represented in the coordinate system $\{x^\mu,w\}$ as
\begin{eqnarray}
  \t{g}_{ab} &=& \t{g}_{AB}dx^A_adx^B_b=(g_{\mu\nu}+\phi^2 A_\mu A_\nu)dx^\mu_adx^\nu_b \nonumber\\
  && +2\phi^2A_\mu dx^\mu_{(a}dw_{b)} +\phi^2dw_adw_b \;, \label{tg_ab1}
\end{eqnarray}
where the parentheses in the indices of a tensor denote symmetrization of the tensor about the indices inside the parentheses. The Einstein summation convention for tensor components is used, i.e., an index appearing in both subscripts and superscripts is summed over all dimensions represented by the index.

The inverse of the $5\times 5$ matrix in equation (\ref{tg_ab}), which is also the component matrix of the inverse five-dimensional metric tensor $\t{g}^{ab}$, is
\begin{eqnarray}
  \t{g}^{AB} = \left(\begin{array}{cc}
    g^{\mu\nu} & -A^\mu\\[1mm]
    -A^\nu & \phi^{-2}+A_\rho A^\rho
  \end{array}\right) \;, \label{tg^ab}
\end{eqnarray}
where the $4\times 4$ matrix $g^{\mu\nu}$ is the inverse of $g_{\mu\nu}$, i.e.,
\begin{eqnarray}
  g_{\mu\nu}g^{\nu\rho}=\delta_\mu^{\;\;\rho} \label{gd_gu}
\end{eqnarray}
where $\delta_\mu^{\;\;\nu}=1$ if $\mu=\nu$ and $0$ otherwise; and
\begin{eqnarray}
  A^\mu\equiv g^{\mu\nu} A_\nu \;. \label{g_A_u}
\end{eqnarray}
Equations (\ref{gd_gu}) and (\ref{g_A_u}) automatically imply
\begin{eqnarray}
  A_\mu=g_{\mu\nu}A^\nu \;. \label{g_A_d}
\end{eqnarray}

By equation (\ref{tg^ab}), the inverse five-dimensional metric tensor is represented as
\begin{eqnarray}
  \t{g}^{ab} &=& g^{\mu\nu}\left(\frac{\partial}{\partial x^\mu}\right)^a\left(\frac{\partial}{\partial x^\nu}\right)^b -2A^\mu\left(\frac{\partial}{\partial x^\mu}\right)^{(a}\left(\frac{\partial}{\partial w}\right)^{b)} \nonumber\\
  && +\left(\frac{1}{\phi^2}+A_\rho A^\rho\right)\left(\frac{\partial}{\partial w}\right)^a\left(\frac{\partial}{\partial w}\right)^b \;. \label{tg^ab1}
\end{eqnarray}
It can be verified that the following reciprocal relation is satisfied
\begin{eqnarray}
  \t{g}_{ab}\t{g}^{bc}=\t{\delta}_a^{\;\;c}\equiv\delta_\mu^{\;\;\nu}dx^\mu_a\left(\frac{\partial}{\partial x^\nu}\right)^c+dw_a\left(\frac{\partial}{\partial w}\right)^c \;, \label{td_ab}
\end{eqnarray}
where $\t{\delta}_a^{\;\;c}$ is the identity map in the five-dimensional spacetime.

In fact, the $5\times 5$ matrices $\t{g}_{AB}$ and $\t{g}^{AB}$ in equations (\ref{tg_ab}) and (\ref{tg^ab}) are inverse to each other if and only if (a) the $4\times 4$ matrices $g^{\mu\nu}$ and $g_{\mu\nu}$ are inverse to each other (eq.~\ref{gd_gu}), and (b) the $A^\mu$ and $A_\mu$ are related by equations (\ref{g_A_u}) and (\ref{g_A_d}). It should be noted that equations (\ref{gd_gu}), (\ref{g_A_u}), and (\ref{g_A_d}) are not independent, since any two of them can give rise to the other.

The KK 4-metric tensor $g_{ab}$ and the electromagnetic potential dual 4-vector $A_a$ are, respectively, a tensor and a vector in the five-dimensional spacetime $(\t{\cal M},\t{g}_{ab})$. The question is how they are expressed in coordinate components in the five-dimensional coordinate system $\{x^\mu,w\}$. Since $g_{\mu\nu}$ are interpreted as the components of the four-dimensional metric in coordinates $\{x^\mu\}$, the $\mu$-$\nu$ components of $g_{ab}$ must be $g_{\mu\nu}$. Then, the general form of $g_{ab}$ must be
\begin{eqnarray}
  g_{ab}=g_{\mu\nu}dx^\mu_a dx^\nu_b+2g_{\mu 4}dx^\mu_{(a}dw_{b)}+g_{44}dw_adw_b \;, \label{gab_w}
\end{eqnarray}
where $g_{\mu 4}$ and $g_{44}$ are to be determined. Similarly, since $A_\mu$ is interpreted as the coordinate component of $A_a$ in $\{x^\mu\}$, and $A^\mu=g^{\mu\nu}A_\nu$ as the coordinate component of $A^a$, we must have
\begin{eqnarray}
  A_a = A_\mu dx^\mu_a+A_4dw_a \label{Aa_w}
\end{eqnarray}
and
\begin{eqnarray}
  A^a=A^\mu\left(\frac{\partial}{\partial x^\mu}\right)^a+A^4\left(\frac{\partial}{\partial w}\right)^a \;, \label{Aa^w}
\end{eqnarray}
where $A_4$ and $A^4$ are to be determined.

By equations (\ref{tg^ab1}) and (\ref{Aa_w}), we have
\begin{eqnarray}
  A^a &=& \t{g}^{ab}A_b = \left(g^{\mu\nu}A_\nu-A^\mu A_4\right)\left(\frac{\partial}{\partial x^\mu}\right)^a \nonumber\\
  && -\left[A_\rho A^\rho-\left(\frac{1}{\phi^2}+A_\rho A^\rho\right)A_4\right]\left(\frac{\partial}{\partial w}\right)^a \;. \hspace{0.3cm} \label{Aa^w1}
\end{eqnarray}
By equation (\ref{g_A_u}), comparison of equation (\ref{Aa^w1}) to equation (\ref{Aa^w}) leads to
\begin{eqnarray}
  A_4=0 \;, \hspace{1cm} A^4=-A_\rho A^\rho \;.
\end{eqnarray}
Thus, we must have
\begin{eqnarray}
  A_a=A_\mu dx^\mu_a \;, \label{Aa_d}
\end{eqnarray}
and
\begin{eqnarray}
  A^a=A^\mu\left(\frac{\partial}{\partial x^\mu}\right)^a-A_\rho A^\rho\left(\frac{\partial}{\partial w}\right)^a \;. \label{Aa_u}
\end{eqnarray}

By equations (\ref{gab_w}) and (\ref{Aa^w}), we have
\begin{eqnarray}
  g_{ab}A^b = \left(A_\mu+g_{\mu 4}A^4\right)dx^\mu_a+\left(g_{\mu 4}A^\mu+g_{44}A^4\right)dw_a \hspace{0.15cm} \label{gab_Aa_x1}
\end{eqnarray}
after submission of equation (\ref{g_A_d}). Since $g_{ab}$ and $A^a$ are interpreted as, respectively, the metric tensor and the electromagnetic potential vector in a four-dimensional spacetime, we must have $g_{ab}A^b=\t{g}_{ab}A^b=A_a$. Then, comparison of equation (\ref{gab_Aa_x1}) to equation (\ref{Aa_d}) leads to
\begin{eqnarray}
  g_{\mu4}=g_{44}=0 \;.
\end{eqnarray}
Thus, in coordinates $\{x^\mu,w\}$ the 4-metric tensor $g_{ab}$ is represented as
\begin{eqnarray}
  g_{ab}=g_{\mu\nu}dx^\mu_a dx^\nu_b \;. \label{gab_exp}
\end{eqnarray}

Then, by $g^{ab}=\t{g}^{ac}\t{g}^{bd}g_{cd}$ we get the inverse 4-metric tensor
\begin{eqnarray}
  g^{ab} &=& g^{\mu\nu}\left(\frac{\partial}{\partial x^\mu}\right)^a\left(\frac{\partial}{\partial x^\nu}\right)^b -2A^\mu\left(\frac{\partial}{\partial x^\mu}\right)^{(a}\left(\frac{\partial}{\partial w}\right)^{b)} \nonumber\\
  && +A^\rho A_\rho\left(\frac{\partial}{\partial w}\right)^a\left(\frac{\partial}{\partial w}\right)^b \;. \label{gab_exp2}
\end{eqnarray}
From equations (\ref{gab_exp}) and (\ref{gab_exp2}) we get
\begin{eqnarray}
  g_a^{\;\;c} &=& g_{ab}g^{bc} =\delta_\mu^{\;\;\nu}dx^\mu_a\left(\frac{\partial}{\partial x^\nu}\right)^c-A_\mu dx^\mu_a\left(\frac{\partial}{\partial w}\right)^c \nonumber\\
  &=& g^c_{\;\;a} \;, \label{gab_exp3}
\end{eqnarray}
just as being expected.

Therefore, given the 4+1 decomposition of the five-dimensional metric in equation (\ref{tg_ab}), the four-dimensional metric tensor $g_{ab}$ and the electromagnetic potential vector $A^a$ are uniquely determined by equations (\ref{gab_exp}) and (\ref{Aa_u}). They are determined by the assumed form of the five-dimensional metric and the self-consistency requirement of the theory, without additional assumptions.

\section{Geometric Interpretation of the Kaluza-Klein Variables}
\label{geom}

For the KK theory to be meaningful, the 4-metric tensor $g_{ab}$ and the electromagnetic potential 4-vector $A_a$ assumed in the KK theory, and the four-dimensional quantities derived from them (e.g., the four-dimensional Ricci tensor $R_{ab}$ and the electromagnetic field antisymmetric tensor $F_{ab}$) must be defined on some four-dimensional manifold---or a four-dimensional submanifold embedded in the five-dimensional manifold $\t{\cal M}$. Such a submanifold ${\cal M}$ should be a hypersurface in $\t{\cal M}$, since $\dim{\cal M}=4=\dim{\cal \t{M}}-1$. Assuming that such a hypersurface has a unit normal $n^a$, which must be spacelike since $({\cal M}, g_{ab})$ is supposed to be a four-dimensional spacetime. That is, all vectors in a vector space tangent to ${\cal M}$ are orthogonal to $n^a$, and $\t{g}_{ab}n^an^b=n^an_a=1$. Then, the 4-metric $g_{ab}$ on ${\cal M}$ must be related to the 5-metric $\t{g}_{ab}$ on $\t{\cal M}$ by $g_{ab}=\t{g}_{ab}-n_an_b$, or, equivalently,
\begin{eqnarray}
  g^{ab}=\t{g}^{ab}-n^an^b \;. \label{gtg_nn}
\end{eqnarray}

The questions are: \textit{does such a hypersurface exist?} If yes, \textit{how is it defined?}

\vspace{0.2cm}

It appears that neither of the above two questions has been seriously considered in the literature, at least to the knowledge of the present author. In his original paper \citep{kal21} (English translation in \cite{app87}, page 61), Kaluza only wrote that ``we are certainly free to consider our space-time to be a four-dimensional part of an $R_5$''. Kaluza used $R_5$ to denote a five-dimensional spacetime. In \citep{kle26a} (English translation in \cite{app87}, page 76), Klein only stated that ``four of the coordinates, $x^1$, $x^2$, $x^3$, $x^4$, say, are to characterize the usual space-time.'' (Klein's $x^1$, $x^2$, $x^3$, $x^4$ are equivalent to our $x^0$, $x^1$, $x^2$, $x^3$ respectively, and his $x^0$ corresponds to our $w$ coordinate.) How is ``a four-dimensional part of an $R_5$'' defined? What is the exact meaning of ``the usual space-time'' in mathematics? These questions have never been clearly answered.

In some references the authors have explicitly identified the four-dimensional spacetime with the hypersurface defined by $w=\mbox{const}$ without any proof or argument. For example, in \citep{ein38} Einstein and Bergmann wrote that ``We consider a four dimensional surface cutting each of the $A$-lines once and only once. We introduce on this surface 4 co\"ordinates $x^a (a= 1 ... 4)$ and assume $x^0$ equal zero on this surface.'' (Section I of \citep{ein38}, subsection ``The Special Co\"ordinate System''). Note that their coordinate $x^0$ corresponds to our $w$, and their $x^a$ correspond to our ${x^\mu} (\mu=0, 1, 2, 3)$. Their ``$A$-lines'' correspond to our $w$-lines. To distinguish it from the electromagnetic potential vector $A^a$ used in this paper, let us denote the 5-vector ``$A$'' used in \citep{ein38} by $\t{A}^a$. In our notations, $\t{A}^a=\phi^{-2}(\partial/\partial w)^a=\phi^{-1}n^a$, which is related to the 4-vector potential $A^a$ by $\t{A}_a=A_a+dw_a$ (see eq.~\ref{Aa_dw} below, Einstein and Bergmann chose $\phi=1$ so that $\t{g}_{ab}\t{A}^a\t{A}^b=1$). Similarly, in \citep{thi48} (English translation in \citep{app87}, page 108), Thiry wrote that ``Kaluza's attempt at a unified theory consists of considering space-time as the $x^0 = \mbox{const.}$ subspace of  a five-dimensional Riemann space, and of assuming this subspace cylindrical with respect to the fifth coordinate $x^0$.'' (Thiry's $x^0$ is equal to our $w$ coordinate.)

There are also people taking different views. For example, \citet{coq90} interpreted the four-dimensional spacetime as a hypersurface orthogonal to the $w$-lines by stating that ``Locally, the 4-dimensional space orthogonal to this vector will be interpreted as the usual space-time''. Their ``this vector'' corresponds to our $n^a=\phi^{-1}(\partial/\partial w)^a$, i.e., the vector $n^a$ in equation (\ref{Aa_dw}) below. However, Coquereaux \& Esposito-Farese did not provide evidence supporting their views. They did not even consider whether ``the 4-dimensional space orthogonal to this vector'' exists or not. As will be shown later in this paper, such a hypersurface does not exist in general.

The view that the submanifold supporting the four-dimensional variables in the KK theory coincides with the hypersurface defined by $w=\mbox{const}$ was disproved in \citep{li16}. Let us use ${\cal S}$ to denote the hypersurface defined by $w=\mbox{const}$, and write its unit normal as $s^a$. The $s^a$ is defined by $\t{g}_{ab}s^as^b=1$ and $\t{g}_{ab}s^a(\partial/\partial x^\mu)^a=0$ for $\mu=0...3$. Thus we must have
\begin{eqnarray}
  s_a=\left(\t{g}^{44}\right)^{-1/2}dw_a = \frac{1}{\sqrt{\phi^{-2}+A_\rho A^\rho}} dw_a \;. \label{s_a}
\end{eqnarray}
By equations (\ref{Aa_u}) and (\ref{gab_exp2}), we have
\begin{eqnarray}
  A^as_a=-\left(\t{g}^{44}\right)^{-1/2}A_\rho A^\rho=-\left(\t{g}^{44}\right)^{-1/2}A_a A^a \;, \hspace{0.3cm} \label{As}
\end{eqnarray}
and
\begin{eqnarray}
  g^{ab}s_b &=& \left(\t{g}^{44}\right)^{-1/2}\left[-A^\mu\left(\frac{\partial}{\partial x^\mu}\right)^a +A^\rho A_\rho\left(\frac{\partial}{\partial w}\right)^a\right] \nonumber\\
  &=& -\left(\t{g}^{44}\right)^{-1/2}A^a \;. \label{gs}
\end{eqnarray}
Hence, the four-dimensional variables $g_{ab}$ and $A^a$ are not orthogonal to $s^a$ unless $A_a=0$.

By equation (\ref{tg_ab1}) we have the metric tensor on ${\cal S}(w=0)$
\begin{eqnarray}
  \hat{g}_{ab} &=& \t{g}_{ab}-s_as_b = \left(g_{\mu\nu}+\phi^2 A_\mu A_\nu\right)dx^\mu_adx^\nu_b \nonumber\\
  && +2\phi^2A_\mu dx^\mu_{(a}dw_{b)} +\frac{\phi^4A_\rho A^\rho}{1+\phi^2A_\rho A^\rho}dw_aw_b \;. \label{hg_ab}
\end{eqnarray}
After restriction of the action of $\hat{g}_{ab}$ on vectors tangent to ${\cal S}$, we get the metric tensor on ${\cal S}$
\begin{eqnarray}
  \hat{g}_{ab}=\left(g_{\mu\nu}+\phi^2 A_\mu A_\nu\right)dx^\mu_adx^\nu_b \;.
\end{eqnarray}
The four-dimensional components of $\hat{g}_{ab}$ are not $g_{\mu\nu}$, but $g_{\mu\nu}+\phi^2 A_\mu A_\nu$. Hence, the KK theory can be treated as \textit{approximately} being defined on ${\cal S}$ only if the electromagnetic field is sufficiently weak so that $|\phi^2A_\mu A_\nu|\ll |g_{\mu\nu}|\sim 1$, i.e., only if
\begin{eqnarray}
  |\phi A_\mu|\ll A_\crit\equiv \frac{c^2}{G^{1/2}}=3.48\times 10^{24}\,\statvolt \;. \label{A_cr}
\end{eqnarray}
If $|\phi A_\mu|\gtrsim A_\crit$, the $g_{ab}$ and $A_a$ are not orthogonal to $s^a$ and the KK theory cannot be treated as being defined on the hypersurface defined by $w=0$.

To find a submanifold ${\cal M}$ supporting the KK four-dimensional variables, we submit equations (\ref{tg^ab1}) and (\ref{gab_exp2}) into equation (\ref{gtg_nn}). We get
\begin{eqnarray}
  n^an^b=\frac{1}{\phi^2}\left(\frac{\partial}{\partial w}\right)^a\left(\frac{\partial}{\partial w}\right)^b \;,
\end{eqnarray}
which immediately leads to a unique solution (up to a sign)
\begin{eqnarray}
  n^a=\phi^{-1}w^a \;, \hspace{0.8cm} w^a\equiv\left(\frac{\partial}{\partial w}\right)^a \;. \label{n^a}
\end{eqnarray}
Thus, if the submanifold ${\cal M}$ exists, its unit normal $n^a$ must be a unit vector tangent to the coordinate lines of the extra dimension. It is easy to verify that both $g_{ab}$ and $A^a$ are orthogonal to $n^a$.

Let us denote the tangent space of the five-dimensional manifold $\t{M}$ at a point $p\in\t{M}$ by $\t{\cal T}_p$, $\dim\t{\cal T}_p=\dim\t{\cal M}=5$. The disjoint union of $\t{\cal T}_p$ at all points of $\t{M}$ is called the tangent bundle of $\t{\cal M}$ and denoted as $\t{\cal T}=\t{\cal T}(\t{\cal M})$. Let ${\cal T}$ be a rank-4 subbundle of $\t{\cal T}$ (called a rank-4 distribution or tangent distribution, or tangent subbundle \citep{lee12}), described by a disjoint union of subspaces containing all vectors and tensors orthogonal to $w^a\propto n^a$ at all points of $\t{\cal M}$, i.e., ${\cal T}=\coprod_{p\in\t{\cal M}}{\cal T}_p$ with $\dim{\cal T}_p=4$. The ${\cal T}$ is a smooth distribution in the sense that for each $p\in\t{\cal M}$ we can find an open neighborhood $\t{\cal O}$ of $p$ such that in $\t{\cal O}$, ${\cal T}$ is spanned by smooth vector and tensor fields orthogonal to $w^a$. We have $g_{ab}$, $g_a^{\;\;b}$, $g^{ab}\in{\cal T}$. The $g_{ab}$ is the metric tensor field in ${\cal T}$. The $g^{ab}$ is the inverse metric tensor field, and $g_a^{\;\;b}$ the identity map in ${\cal T}$.\footnote{The $g_a^{\;\;b}=\t{g}_a^{\;\;b}-n_an^b$ is also the projection operator mapping a vector (and a tensor) in $\t{\cal T}$ to a vector (and a tensor) in ${\cal T}$.} Since $A_aw^a=0=A^aw_a$, we have $A_a$, $A^a\in{\cal T}$, too.

By equations (\ref{tg_ab1}) and (\ref{Aa_d}), we have
\begin{eqnarray}
  \phi^2A_a=\phi^2A_\mu dx^\mu_a=\t{g}_{bc}\left(\frac{\partial}{\partial x^\mu}\right)^b\left(\frac{\partial}{\partial w}\right)^cdx^\mu_a \;.
\end{eqnarray}
By equation (\ref{td_ab}), we have
\begin{eqnarray}
  dx^\mu_a\left(\frac{\partial}{\partial x^\mu}\right)^b=\t{\delta}_a^{\;\;b}-dw_a\left(\frac{\partial}{\partial w}\right)^b \;.
\end{eqnarray}
Hence, we get
\begin{eqnarray}
  \phi^2A_a &=& \t{g}_{bc}\left(\frac{\partial}{\partial w}\right)^c\left[\t{\delta}_a^{\;\;b}-dw_a\left(\frac{\partial}{\partial w}\right)^b\right] \nonumber\\
  &=& \t{g}_{ac}\left(\frac{\partial}{\partial w}\right)^c-\phi^2 dw_a \;, \label{fi2_Aa}
\end{eqnarray}
where we have used $\t{g}_{bc}(\partial/\partial w)^b(\partial/\partial w)^c=\t{g}_{44}=\phi^2$. Submitting the definition of $n^a$ in equation (\ref{n^a}) into equation (\ref{fi2_Aa}), we get the relation
\begin{eqnarray}
  A_a=\phi^{-1}n_a-dw_a \;. \label{Aa_dw}
\end{eqnarray}

Equation (\ref{Aa_dw}) states that $A^a$ is obtained from orthogonal decomposition of the vector field $\mu^a\equiv\t{g}^{ab}dw_b=(g^{44})^{1/2} s^a$: one component is along the direction of $w^a$ (i.e., the direction of $n^a$ by eq.~\ref{n^a}), the other (which is $\propto -A^a$) is in the direction orthogonal to $w^a$.

By $\t{\nabla}_{[a}dw_{b]}=0$, where the brackets in the indices of a tensor denote antisymmetrization of the tensor about the indices inside the brackets, equation (\ref{Aa_dw}) implies that $\t{\nabla}_{[a}A_{b]} = \t{\nabla}_{[a}\left(\phi^{-1}n_{b]}\right)$, i.e.,
\begin{eqnarray}
  \t{\nabla}_{[a}A_{b]} = n_{[b}\t{\nabla}_{a]}\phi^{-1}+\phi^{-1}\t{\nabla}_{[a}n_{b]} \;. \label{dAa_dw}
\end{eqnarray}
Although $\t{\nabla}_a$ is the derivative operator associated with the metric $\t{g}_{ab}$, equation (\ref{dAa_dw}) remains valid if $\t{\nabla}_a$ is replaced by any derivative operator. By equation (\ref{dAa_dw}), we have the antisymmetric tensor of the electromagnetic field
\begin{eqnarray}
  F_{ab} \equiv 2g_a^{\;\;c}g_b^{\;\;d}\t{\nabla}_{[c}A_{d]}= 2\phi^{-1}g_a^{\;\;c}g_b^{\;\;d}\t{\nabla}_{[c}n_{d]} \;, \label{F_dn}
\end{eqnarray}
where we have used the identity $g_a^{\;\;b}n_b=0$. By definition, $F_{ab}\in{\cal T}$ since $F_{ab}w^b=0=F_{ab}w^a$.

If $w^a$ were timelike and $n_an^a=-1$, the quantity
\begin{eqnarray}
  \omega_{ab}\equiv g_a^{\;\;c}g_b^{\;\;d}\t{\nabla}_{[d}n_{c]} \label{vot_ab}
\end{eqnarray}
would be called the vorticity tensor of the congruence of the integral curves of the vector field $w^a$ \citep{haw73}. Similarly, the expansion tensor would be defined by
\begin{eqnarray}
  \theta_{ab}\equiv g_a^{\;\;c}g_b^{\;\;d}\t{\nabla}_{(c}n_{d)}=\frac{1}{2}\t{\pounds}_ng_{ab} \;, \label{exp_ab}
\end{eqnarray}
where $\t{\pounds}_n$ is the Lie derivative generated by $n^a$ (as usual, the tilde above $\pounds_n$ indicates that the operation is in the five-dimensional spacetime). Now $w^a$ is spacelike and $n_an^a=1$. We can still define the $\omega_{ab}$ and $\theta_{ab}$ by equations (\ref{vot_ab}) and (\ref{exp_ab}) and call them the vorticity tensor and the expansion tensor, respectively, of the integral curves of $w^a$. Both $\omega_{ab}$ and $\theta_{ab}$ are $\in{\cal T}$, since they both are orthogonal to $w^a$. Then, equation (\ref{F_dn}) is equivalent to
\begin{eqnarray}
  F_{ab}=-2\phi^{-1}\omega_{ab} \;. \label{F_om}
\end{eqnarray}
Therefore, \textit{the electromagnetic field antisymmetric tensor in the KK theory is proportional to the vorticity tensor of the congruence of the extra dimension coordinate curves.}

\section{Submanifold Supporting the Effective Four-Dimensional Theory does not Exist}
\label{subm}

As shown in section \ref{geom}, the KK four-dimensional field variables $g_{ab}$ and $A_a$ are contained in a rank-4 distribution ${\cal T}$ with members orthogonal to the extra dimension coordinate vector field $w^a$. In order that there exists an integral submanifold of ${\cal T}$, i.e., a submanifold ${\cal M}\subseteq\t{\cal M}$ whose tangent space at any $p\in{\cal M}$ coincides with ${\cal T}_p$, the distribution ${\cal T}$ must be involutive, i.e., for any $X^a,Y^a\in{\cal T}$ the commutator (Lie bracket) $[X,Y]^a\in{\cal T}$ (Frobenius's theorem, \citep{lee12}). The condition that the tangent subbundle ${\cal T}$ is integrable is equivalent to that the vector field $w^a$ is hypersurface orthogonal, i.e., there exists a hypersurface orthogonal to $w^a$.

By Frobenius's theorem, $w^a$ is hypersurface orthogonal if and only if $[X,Y]^a\in{\cal T}$ for all $X^a,Y^a\in{\cal T}$, which is mathematically equivalent to the condition that there is a dual vector $v_a$ such that \citep{wal84,str13}
\begin{eqnarray}
  \t{\nabla}_{[a}w_{b]}=w_{[a}v_{b]} \;. \label{frob1}
\end{eqnarray}

Equation (\ref{frob1}) is equivalent to the condition that the congruence of the spacelike curves tangent to $w^a$ is vorticity-free, which is proved as follows. Since $w^a=\phi n^a$, equation (\ref{frob1}) is equivalent to
\begin{eqnarray}
  \t{\nabla}_{[a}n_{b]}=n_{[a}v^\prime_{b]} \;, \label{frob2}
\end{eqnarray}
where $v^\prime_a=v_a+\t{\nabla}_a\ln\phi$. Then we have
\begin{eqnarray}
  g_a^{\;\;c}g_b^{\;\;d}\t{\nabla}_{[c}n_{d]}=g_a^{\;\;c}g_b^{\;\;d}n_{[c}v^\prime_{d]}=0 \;, 
\end{eqnarray}
since $g_a^{\;\;c}n_c=0$. By equation (\ref{vot_ab}) we get then
\begin{eqnarray}
  \omega_{ab}=0 \;. \label{frob3}
\end{eqnarray}

Equation (\ref{frob2}) can also be derived from equation (\ref{frob3}), since by $ \t{\nabla}_an_b =\theta_{ab}+\omega_{ba}+n_aa_b$
we have
\begin{eqnarray}
  \t{\nabla}_{[a}n_{b]}=\omega_{ba}+n_{[a}a_{b]} \;,
\end{eqnarray}
where $a_a\equiv n^b\t{\nabla}_bn_a$. Thus, $\omega_{ab}=0$ implies $\t{\nabla}_{[a}n_{b]}=n_{[a}a_{b]}$, i.e., equation (\ref{frob2}) if we take $v^\prime_b=a_b$. Therefore, \textit{a necessary and sufficient condition for a congruence of timelike or spacelike curves to be hypersurface orthogonal is that the congruence is vorticity-free.}

Only if $w^a$ is hypersurface orthogonal the distribution ${\cal T}$ in a neighborhood of any point is integrable and can be spanned by coordinate base vector fields \citep{wal84}. By equations (\ref{F_om}) and (\ref{frob3}), $w^a$ is hypersurface orthogonal if and only if
\begin{eqnarray}
  F_{ab}=0 \;. \label{frob4}
\end{eqnarray}
Thus, \textit{the submanifold ${\cal M}$ supporting the KK variables exists if and only if the electromagnetic field antisymmetric tensor vanishes.} This restriction on the KK theory is too strong, as there can be no unification of the gravitational and electromagnetic interactions if the electromagnetic field vanishes.

\section{Action and Compactification of the Extra Dimension}
\label{action}

When the cylinder condition is satisfied, i.e., all components of the five-dimensional metric tensor are independent of the extra dimension coordinate \citep{kal21}, the Maxwell equations can easily be derived from the action principle.\footnote{If the cylinder condition is dropped the derived field equations are much more complicated, see, e.g., \citep{ove97}.} The cylinder condition is equivalent to the requirement that $w^a$ is a Killing vector of the five-dimensional spacetime, i.e.,
\begin{eqnarray}
  \t{\pounds}_w\t{g}_{ab} = 0 \;.  \label{kill_eq}
\end{eqnarray}
When the above condition is satisfied, the Ricci scalar $\t{R}$ associated with the five-dimensional bulk metric $\t{g}_{ab}$ is related to the Ricci scalar $R$ associated with the KK four-dimensional metric $g_{ab}$ by \citep{bai87,coq90,wil15}
\begin{eqnarray}
  \t{R}=R-\frac{\phi^2}{4}F_{ab}F^{ab}+\t{\nabla}_av^a \;, \label{tR_R_FF}
\end{eqnarray}
where $v^a$ is a vector.

The determinant of the five-dimensional metric, $\t{g}=\det\t{g}_{AB}$, is related to the determinant of the four-dimensional metric, $g=\det g_{\mu\nu}$, by $\t{g}=\phi^2 g$. Thus, we have $\sqrt{-\t{g}}=\phi\sqrt{-g}$ and the five-dimensional action of gravity
\begin{eqnarray}
  I_g &=& \frac{1}{\t{G}}\int\t{R}\sqrt{-\t{g}}\,d^4xdw \nonumber\\
  &=& \frac{1}{\t{G}}\int dw\int\phi\left(R-\frac{\phi^2}{4}F_{ab}F^{ab}\right)\sqrt{-g}\,d^4x \;, \hspace{0.4cm}\label{act1}
\end{eqnarray}
where $\t{G}$ is the five-dimensional gravitational constant. The divergence term in equation (\ref{tR_R_FF}), $\t{\nabla}_av^a$, has been dropped off since it has no contribution to the action integral under appropriate boundary conditions.

\begin{figure}[ht]
\vspace{3pt}
\begin{center}\includegraphics[angle=0,scale=0.6]{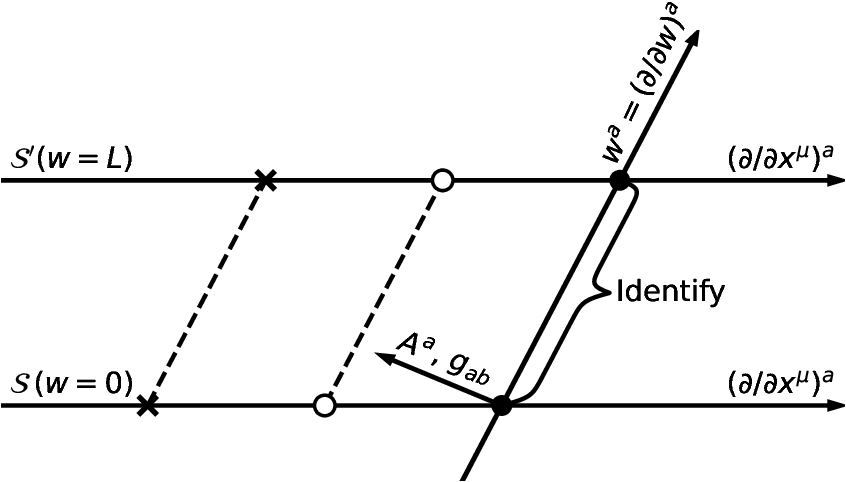}\end{center}
\caption{Under the cylinder condition the five-dimensional spacetime can be compactified along the direction of extra dimension [i.e., the direction of $w^a=(\partial/\partial w)^a$]. This way, the hypersurface ${\cal S}(w=0)$ in $(\tilde{M}, \t{g}_{ab})$ is identified with the ${\cal S}^\prime(w=L)$ under the map generated by $w^a$. The black dot on ${\cal S}$ is identified with the black dot on ${\cal S}^\prime$, the circle on ${\cal S}$ identified with the circle on ${\cal S}^\prime$, and so on (as indicated by dashed lines). The KK variables $A^a$ and $g_{ab}$ are orthogonal to $w^a$ hence not tangent to ${\cal S}$, since in general $w^a$ is not orthogonal to ${\cal S}$. In fact, $w^a$ is not orthogonal to any hypersurface unless the electromagnetic field $F_{ab}$ vanishes. Note that to make the extra dimension unaccessible to current experiments its circumference $C_w=\phi L$ has to be small.
}
\label{comp}
\end{figure}

Since $w^a$ is a Killing vector field, the five-dimensional spacetime can be compactified along the direction of extra dimension, i.e., the direction of $w$-lines (Fig.~\ref{comp}). That is, a spacetime point $\{x^\mu, w\}$ in $(\tilde{M},\t{g}_{ab})$ is identified with the spacetime point $\{x^\mu, w+L\}$ in $(\tilde{M},\t{g}_{ab})$, where $L>0$ is a constant. Then, the value of $w$ is restricted to the region of $[0,L)$, which leads to $\int dw=L$. To have the circumference of the extra dimension---which is $C_w\equiv\phi\int dw =\phi L$---to be constant, $\phi$ must be constant \citep{bai87}.\footnote{It should be noted that ${\phi}$ cannot be constant if the vacuum Einstein field equation $\t{R}_{ab}=0$ is imposed. This follows because the equation $\t{R}_{ww}=0$ entails that $$ \t{\nabla}^a\t{\nabla}_a\phi = \frac{\phi^3}{4}F^{ab}F_{ab} \;, $$ as first noted by Jordan \cite{jor47} and Thiry \cite{thi48}.} For the extra dimension to be unaccessible to current experiments the length $C_w=\phi L$ must be sufficiently small \citep{kle26a}. Then, equation (\ref{act1}) becomes
\begin{eqnarray}
  I_g = \frac{C_w}{\t{G}}\int\left(R-\frac{\phi^2}{4}F_{ab}F^{ab}\right)\sqrt{-g}\,d^4x \;. \label{act2}
\end{eqnarray}

The appearance of the term $-(\phi^2/4)F_{ab}F^{ab}$ in the Lagrangian density in equation (\ref{act2}) guarantees that the Maxwell equations can be derived from the five-dimensional Einstein field equation by variation of the action $I_g$ with respect to the potential vector $A^a$. In fact, if we identify $\t{G}/C_w$ as the four-dimensional gravitational constant $G=1$ (i.e., $C_w=\t{G}$) and $\phi = 2$, the action in equation (\ref{act2}) reduces to the total action of gravity and electromagnetic fields in a four-dimensional spacetime (see \citep{wal84}, Appendix E). Note that, however, all the quantities ($F_{ab}$, $R$, and $g$) appearing in the integral of equation (\ref{act2}) are defined in the tangent subbundle ${\cal T}$ orthogonal to $w^a$, since they all are derived from $A_a$, $g_{ab}$, and the derivative operator $\nabla_a$ associated with $g_{ab}$.

According to the results in section \ref{subm}, the rank-4 distribution ${\cal T}$ is not integrable and hence $g_{ab}$, $A_a$, and quantities derived from them are not tangent to any four-dimensional submanifold embedded in $\t{\cal M}$ unless the electromagnetic field $F_{ab}$ vanishes. Although under the cylinder condition the four-dimensional Einstein field equation and the Maxwell equations are successfully derived from the five-dimensional Einstein field equation through the action principle, these equations are not supported by a four-dimensional submanifold hence do not define a four-dimensional spacetime.

\begin{figure}[ht]
\vspace{3pt}
\begin{center}\includegraphics[angle=0,scale=0.6]{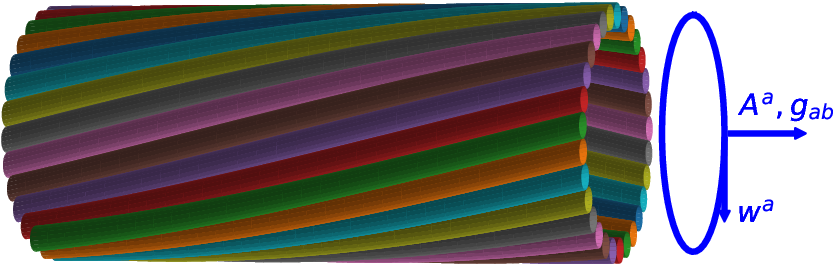}\end{center}
\caption{The Kaluza-Klein tube obtained by compactification of the extra dimension in the direction of $w^a$. The five-dimensional spacetime tube is made of twisted wires, each wire representing a hypersurface defined by $w=\mbox{const}$. The blue circle on the right represents the $w$-coordinate line with a length of $C_w=\phi L$. The KK variables $A^a$ and $g_{ab}$ are contained in a four-dimensional tangent subbundle ${\cal T}$ orthogonal to $w^a$.
}
\label{rope}
\end{figure}

The manifold structure of the KK theory, after the extra dimension is compactified, is depicted in Fig.~\ref{rope}. The five-dimensional spacetime ``tube'' is made of twisted four-dimensional ``wires'', with each wire representing a hypersurface $w=\mbox{const}$. The transverse cross-section of the spacetime tube corresponds to the $w$-coordinate lines, i.e., curves whose tangent vectors are $w^a=(\partial/\partial w)^a$, as indicated by the blue circle in the figure. The KK variables $A^a$, $g_{ab}$, and the associated distribution ${\cal T}$, are in the longitudinal direction along the tube (i.e., the direction perpendicular to $w^a$). They are not tangent to any four-dimensional submanifold. Thus, the action in equation (\ref{act2}) is defined in the rank-4 tangent subbundle or distribution ${\cal T}$, but not defined on a four-dimensional submanifold.

\section{Summary and Discussion}
\label{disc}

All existing physical theories are defined on a smooth manifold with or without a well-defined spacetime metric. In the KK theory, the five-dimensional theory is defined on a five-dimensional manifold with a Lorentz metric determined by the five-dimensional Einstein field equation. The four-dimensional metric tensor and the electromagnetic potential vector assumed in the KK theory must be  defined on a four-dimensional submanifold (i.e., a hypersurface) embedded in the five-dimensional manifold, in order for the derived four-dimensional theory (including the four-dimensional Einstein field equation and the Maxwell equations) to be able to describe the four-dimensional world where we live and do physical experiments. But this is not the case, as has been shown in this paper.

In general, the four-dimensional KK variables $g_{ab}$, $A_a$, and other geometric quantities derived from them (e.g., the four-dimensional Ricci tensor $R_{ab}$ and the electromagnetic field antisymmetric tensor $F_{ab}$) are in a four-dimensional subbundle that is not tangent to any four-dimensional submanifold, since by the KK construction $g_{ab}$ and $A_a$ are orthogonal to the vector field $w^a$ generating the extra dimension but $w^a$ is not hypersurface orthogonal unless the electromagnetic field vanishes. Thus, the results presented in the paper lead us to such a paradox: the KK theory is valid mathematically only if the electromagnetic field derived from the KK theory vanishes. This is a general conclusion, independent of the cylinder condition adopted for derivation of the four-dimensional field equations.

When the electromagnetic field is weak and has a negligible effect on the spacetime metric, i.e., when condition (\ref{A_cr}) is satisfied, the four-dimensional metric tensor and the electromagnetic potential vector can be regarded as \textit{approximately} being defined on the hypersurface of $w=\mbox{const}$. But then the KK theory becomes an approximate and weak-field limit theory, conflicting the original spirit of unification of gravitational and electromagnetic interactions. In addition, without a precisely defined four-dimensional submanifold supporting the four-dimensional variables, it is hard to accept the approximate theory since it is not well defined in mathematics. An ultimate solution to the problem raised in this paper may be given by a different 4+1 decomposition of a five-dimensional spacetime metric as having been proposed in \citep{li16}, where the four-dimensional spacetime is defined on a hypersurface that is not orthogonal to the extra dimension, but then the theory is different from the KK theory since an electromagnetic field equation with a curvature-coupled term is derived.


\begin{acknowledgments}
The author thanks the anonymous adjudicator for a very enlightening report, which has stimulated the author to think more deeply and more widely about the problem discussed in the paper. The report has also helped to improve the presentation of the paper. This work was supported by the NSFC grants program (no.~11973014).
\end{acknowledgments}



\begin{thebibliography}{99}
  
\bibitem[Kaluza(1921)]{kal21}
  T. Kaluza, \textit{Zum Unit\"atsproblem der Physik}, Sitzungsber. Preuss. Akad. Wiss., 966 (1921)
\bibitem[Klein(1926a)]{kle26a}
  O. Klein, \textit{Quantentheorie und F\"unfdimensionale Relativit\"atstheorie}, Z. Phys. {\bf 37}, 895 (1926)
\bibitem[Klein(1926b)]{kle26b}
  O. Klein, \textit{The Atomicity of Electricity as a Quantum Theory Law}, Nature {\bf 118}, 516 (1926)
\bibitem[Nordstr{\"o}m(1914)]{nor14}
  G. Nordstr{\"o}m, \textit{\"Uber die Moglichkeit, das Electromagnetische Feld und das Gravitationsfeld zu Vereinigen}, Phys. Z. {\bf 15}, 504 (1914)
\bibitem[Cremmer \& Scherk(1976)]{cre76}
  E. Cremmer \& J. Scherk, \textit{Spontaneous Compactification of Space in an Einstein-Yang-Mills-Higgs Model}, Nucl. Phys. B {\bf 108}, 409 (1976)
\bibitem[Cremmer \& Scherk(1977)]{cre77}
  E. Cremmer \& J. Scherk, \textit{Spontaneous Compactification of Extra Space Dimensions}, Nucl. Phys. B {\bf 118}, 61 (1977)
\bibitem[Luciani(1978)]{luc78}
  J. F. Luciani, \textit{Space-time Geometry and Symmetry Breaking}, Nucl. Phys. B {\bf 135}, 111 (1978)
\bibitem[Englert(1982)]{eng82}
  F. Englert, \textit{Spontaneous Compactification of Eleven-Dimensional Supergravity}, Phys. Lett. B {\bf 119}, 339 (1982)
\bibitem[Li \& Gott(1998)]{li98}
  L.-X. Li \& J. R. Gott, \textit{Inflation in Kaluza-Klein Theory: Relation between the Fine-Structure Constant and the Cosmological Constant}, Phys. Rev. D {\bf 58}, 103513 (1998)
\bibitem[van Nieuwenhuizen(1981)]{nie81}
  P. van Nieuwenhuizen, \textit{Supergravity}, Phys. Rep. {\bf 68}, 189 (1981)
\bibitem[Duff, Nilsson, \& Pope(1986)]{duf86}
  M. J. Duff, B. E. W. Nilsson, \& C. N. Pope, \textit{Kaluza-Klein Supergravity}, Phys. Rep. {\bf 130}, 1 (1986)
\bibitem[Green, Schwarz, \& Witten(2012)]{gre12}
  M. B. Green, J. H. Schwarz, \& E. Witten, \textit{Superstring Theory, 25th Anniversary Edition, Vols. I and II} (Cambridge University Press, Cambridge, 2012)
\bibitem[Polchinski(2003)]{pol03}
  J. Polchinski, \textit{String Theory, Vols. I and II} (Cambridge University Press, Cambridge, 2003)
\bibitem[Becker, Becker, \& Schwarz(2007)]{bec07}
  K. Becker, M. Becker, \& J. H. Schwarz, \textit{String Theory and M-Theory: A Modern Introduction} (Cambridge University Press, Cambridge, 2007)
\bibitem[Kiritsis(2019)]{kir19}
  E. Kiritsis, \textit{String Theory in a Nutshell, 2nd edition} (Princeton University Press, Princeton, 2019)
\bibitem[Randall \& Sundrum(1999a)]{ran99a}
  L. Randall \& R. Sundrum, \textit{Large Mass Hierarchy from a Small Extra Dimension}, Phys. Rev. Lett. {\bf 83}, 3370 (1999)
\bibitem[Randall \& Sundrum(1999)b]{ran99b}
  L. Randall \& R. Sundrum, \textit{An Alternative to Compactification}, Phys. Rev. Lett. {\bf 83}, 4690 (1999)
\bibitem[Maartens \& Koyama(2010)]{maa10}
  R. Maartens \& K. Koyama, \textit{Brane-World Gravity}, Living Rev. Relativity {\bf 13}, 5 (2010)
\bibitem[Nath \& Yamaguchi(1999)]{nat99}
  P. Nath \& M. Yamaguchi, \textit{Probing the Nature of Compactification with Kaluza-Klein Excitations at the Large Hadron Collider}, Phys. Lett. B {\bf 466}, 100 (1999)
\bibitem[Bhattacharyya et al.(2009)]{bha09}
  G. Bhattacharyya, A. Datta, S. K. Majee, \& A. Raychaudhuri, \textit{Exploring the Universal Extra Dimension at the LHC}, Nucl. Phys. B {\bf 821}, 48 (2009)
\bibitem[Deutschmann et al.(2017)]{deu17}
  N. Deutschmann, T. Flacke, \& J. S. Kim, \textit{Current LHC Constraints on Minimal Universal Extra Dimensions}, Phys. Lett. B {\bf 771}, 515 (2017)
\bibitem[Wald(1984)]{wal84}
  R. M. Wald, \textit{General Relativity} (The University of Chicago Press, Chicago, 1984)
\bibitem[de~Witt(1964)]{dew64}
  B. S. de~Witt, \textit{Dynamical Theory of Groups and Fields}, in \textit{Relativity Groups and Topology, Les Houches 1963} (Gordon and Breach Science Publishers, 1964), p. 585
\bibitem[Kerner(1987)]{ker68}
  R. Kerner, \textit{Generalization of the Kaluza-Klein Theory for an Arbitrary non-Abelian Gauge Group}, Ann. Inst. Henri Poincar{\'e} {\bf 9}, 143 (1968)
\bibitem[Cho \& Freund(1975)]{cho75}
   Y. M. Cho \& P. G. O. Freund, \textit{Non-Abelian gauge fields as Nambu-Goldstone fields}, Phys. Rev. D {\bf 12}, 1711 (1975)
\bibitem[Appelquist et al.(1987)]{app87}
  T. Appelquist, A. Chodos, \& P. G. O. Freund (editors), \textit{Modern Kaluza-Klein Theories} (Addison-Wesley, Amsterdam, 1987)
\bibitem[Einstein \& Bergmann(1938)]{ein38}
  A. Einstein \& P. Bergmann, \textit{On a Generalization of Kaluza's Theory of Electricity}, Ann. Math. {\bf 39}, 683 (1938)
\bibitem[Thiry(1948)]{thi48}
  M. Y. Thiry, \textit{Les \'Equations de la Th{\'e}orie Unitaire de Kaluza}, Compt. Rend. Acad. Sci. Paris. {\bf 226}, 216 (1948)
\bibitem[Coquereaux \& Esposito-Farese(1990)]{coq90}
  R. Coquereaux \& G. Esposito-Farese, \textit{The Theory of Kaluza-Klein-Jordan-Thiry Revisited}, Ann. Inst. Henri Poincar{\'e} {\bf 52}, 113 (1990)
\bibitem[Li(2016)]{li16}
  L.-X. Li, \textit{A New Unified Theory of Electromagnetic and Gravitational Interactions}, Front. Phys. {\bf 11}, 110402 (2016)
\bibitem[Lee(2012)]{lee12}
  J. M. Lee, \textit{Introduction to Smooth Manifolds, 2nd edition} (Springer, Berlin, 2012)
\bibitem[Hawking \& Ellis(1973)]{haw73}
  S. W. Hawking \& G. F. R. Ellis, \textit{The Large Scale Structure of Space-Time} (Cambridge University Press, Cambridge, 1973)
\bibitem[Bailin \& Love(1987)]{bai87}
  D. Bailin \& A. Love, \textit{Kaluza-Klein Theories}, Rep. Prog. Phys. {\bf 50}, 1087 (1987)
\bibitem[Williams(2015)]{wil15}
  L. L. Williams, \textit{Field Equations and Lagrangian for the Kaluza Metric Evaluated with Tensor Algebra Software}, Journal of Gravity {\bf 2015}, 901870 (2015)
\bibitem[Straumann(2013)]{str13}
  N. Straumann, \textit{General Relativity, 2nd edition} (Springer, Berlin, 2013)
\bibitem[Overduin \& Wesson(1997)]{ove97}
  J. M. Overduin \& P. S. Wesson, \textit{Kaluza-Klein Gravity}, Phys. Rep. {\bf 283}, 303 (1997)
\bibitem[Jordan(1947)]{jor47}
  P. Jordan, \textit{Erweiterung der projektiven Relativit\"atstheo-rie}, Ann. Physik {\bf 436}, 219 (1947)
  
\end{thebibliography}
\end{document}